# Retention Time Prediction for Chromatographic Enantioseparation by Quantile Geometry-enhanced Graph Neural Network

Hao Xu[a,b], Jinglong Lin[a], Dongxiao Zhang[c,d,*], Fanyang Mo[a,*]

[a] *School of Materials Science and Engineering, Peking University, Beijing, 100871, P. R. China.*

[b] *BIC-ESAT, ERE, and SKLTCS, College of Engineering, Peking University, Beijing, 100871, P. R. China.*

[c] *Department of Mathematics and Theories, Peng Cheng Laboratory, Shenzhen, 518000, P. R. China.*

[d] *National Center for Applied Mathematics Shenzhen (NCAMS), Southern University of Science and Technology, Shenzhen, 518000, P. R. China.*

[*] Corresponding author
[*] Correspondence: zhangdx@sustech.edu.cn (D. Z.), fmo@pku.edu.cn (F. M.)

**Abstract**
A new research framework is proposed to incorporate machine learning techniques into the field of experimental chemistry to facilitate chromatographic enantioseparation. A documentary dataset of chiral molecular retention times (CMRT dataset) in high-performance liquid chromatography is established to handle the challenge of data acquisition. Based on the CMRT dataset, a quantile geometry-enhanced graph neural network is proposed to learn the molecular structure-retention time relationship, which shows a satisfactory predictive ability for enantiomers. The domain knowledge of chromatography is incorporated into the machine learning model to achieve multi-column prediction, which paves the way for chromatographic enantioseparation prediction by calculating the separation probability. Experiments confirm that the proposed research framework works well in retention time prediction and chromatographic enantioseparation facilitation, which sheds light on the application of machine learning techniques to the experimental scene and improves the efficiency of experimenters to speed up scientific discovery.

**Introduction**
In recent years, the rapid development of machine learning intelligence has brought prosperity to the field of 'machine learning for chemistry'[1], which spawns a series of applications including molecular properties prediction[2], drug discovery[3], and retrosynthetic analysis[4–6]. Although diversified machine learning models have been invented to accomplish requirements in many research scenarios[7–9], fundamental limitations still lie in the aspects of dataset generation and molecular representations, which hinder the integration of machine learning and chemistry.

Datasets are fundamental to machine learning since the quantity and quality of data directly relate to the performance of machine learning models. Unfortunately, the generation of chemical data is usually time-consuming and labor-intensive due to the experimental attributes in chemistry. Therefore, high-throughput techniques combined with automation have been developed to accumulate standardized experimental data efficiently[10,11]. However, high-throughput systems are usually expensive and targeted at specific scenarios, which is difficult to promote to broader fields. An alternative way is collecting data from published articles, but the quality usually varies from objective factors. It means that the uncertainty of data needs to be taken into consideration.

Molecular representation is another issue that needs to be handled properly. Chemical molecules usually have a variety of classic representation ways, including SMILES[12], fingerprints[13], and descriptors[14]. Although these ways have achieved gratifying performance in constructing quantitative structure-activity relationships (QSAR), they have difficulty in representing 3D conformer-related properties like charity (Fig. 1a), which confines their further application. Fortunately, the derivatives of graph neural network (GNN), including geometry-enhanced graph neural network (GeoGNN)[15] and Uni-mol[16], attempted to incorporate 3D information to enhance the molecular graph representation. However, massive data are required for training, which is unaffordable in the experimental scene where data are usually scarce and expensive.

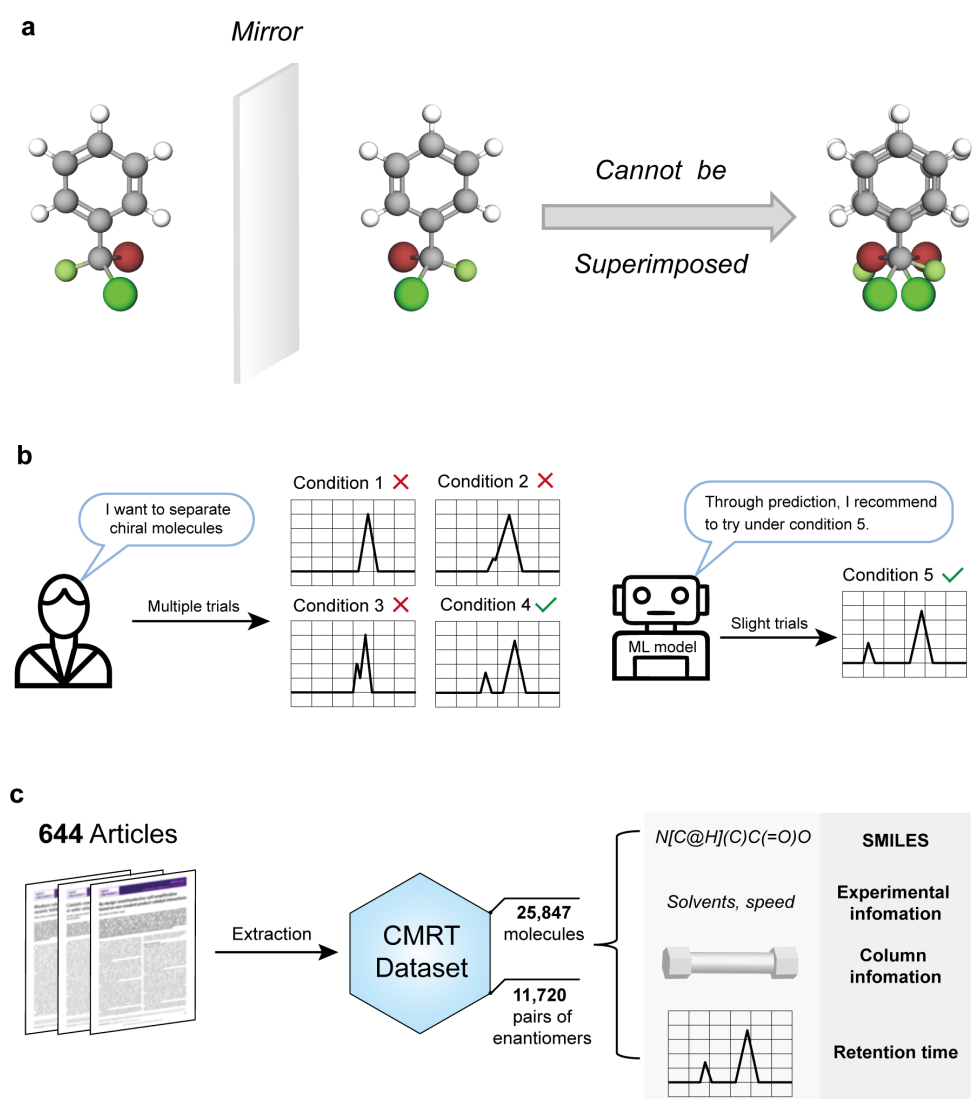

**Fig. 1. The scheme for chromatographic enantioseparation. a,** The diagram for chiral molecules, which are mirror images of each other, but not superimposable. **b,** The comparison between classic separation ways by multiple trials with different conditions and machine learning (ML) models that can recommend the most suitable conditions with the highest separation probability. **c,** The generation procedure and contents of the chiral molecular retention time dataset (CMRT dataset) in this work.

In view of the above-mentioned pain points, we explore a research framework to incorporate machine learning techniques into practical problems in experimental chemistry. The prediction for chromatographic enantioseparation is presented as a persuasive example in this work, which is of great significance in synthetic chemistry, material science, and biopharmaceutical[17–19]. High-performance liquid chromatography (HPLC) is the mainstream way for chromatographic enantioseparation[20], however, the choice of the experimental condition needs trial-and-error, which is tedious and time-consuming since each trial may take tens of minutes (Fig. 1b). Therefore, in this work, we construct a machine learning model that can predict the retention time (RT) of given chiral molecules and recommend the most suitable condition with the highest possibility of separation (Fig. 1b). Different from previous works that focus merely on the quantitative structure–retention relationship (QSRR) models[21–24], we take a step further to consider how the machine learning models can promote chemical experiments practically. To this end, we pay more attention to the normal-phase HPLC that is usually employed to separate chiral molecules instead of reversed-phase HPLC in existing literatures[24,25]. Our contribution can be summarized as follows:

1. A chiral molecular retention time dataset (CMRT dataset) is established in this work by collecting experimental data reported in 644 articles about asymmetric catalysis. The CMRT dataset constitutes the retention time of 25,847 molecules, which contains 11,720 pairs of enantiomers, experimental information, and HPLC column information. The molecules are recorded in the form of SMILES (Fig. 1c).

2. A machine learning framework called quantile geometry-enhanced graph neural network (QGeoGNN) is constructed by combining quantile learning and GeoGNN, which takes the data uncertainty and chiral molecular representation into consideration and shows satisfactory performance in predicting retention times of chiral molecules. In this framework, the domain knowledge of chromatography and experimental conditions are also incorporated into the model to enhance its extendibility.

3. The prediction model is able to instruct chromatographic enantioseparation by predicting the separation possibility in different conditions, and thus eliminating repeated trials, which provides a framework for the utilization of machine learning techniques to facilitate the experimental chemistry where data are expensive and unstandardized.

## Results
### Backgrounds and the CMRT dataset
As a ubiquitous phenomenon in nature, molecular chirality is a significant factor that affects molecular properties. A pair of chiral molecules is termed as enantiomers, which are mirror images of each other, but not superimposable (Fig. 1a). Although the molecular constitution of enantiomers is identical with the same atoms and bonds, their properties may be disparate due to the chirality. As an example, left-handed thalidomide is an effective tranquilizer for parturition, while the right-handed enantiomer leads to developmental abnormality in fetuses, and the mixture of enantiomers in the drug once triggered a tragedy[26]. Generally, separating chiral molecules is a challenge because the constitutions of chiral molecules are identical. In order to obtain enantiomerically pure compounds, several chromatographic enantioseparation techniques have been developed in the past decades to separate and analyze the chiral compounds[27]. Among these techniques, high-performance liquid chromatography (HPLC) becomes the mainstream way benefiting from its high efficiency and popularity[20]. In HPLC, the retention time (RT) is a fundamental characteristic, which is defined as the time of chromatographic components from injection to peak (Fig. 1b). It can be used as a qualitative basis for chromatographic enantioseparation since each compound corresponds to a retention time under certain condition. In the chromatographic enantioseparation, the normal-phase HPLC column is adopted where stereoregular chiral polymers are employed as chiral stationary phases (CSPs) to differentiate chiral molecules. Considering that there exist diversified CSPs, different types of HPLC columns have discrepant chiral recognition capacities for different chiral compounds. However, the choice of experimental conditions, including the column type, flow speed, and elution proportion, is currently determined by experience and repeated trials (Fig. 1b). Therefore, this work attempts to construct a machine learning prediction model to predict retention times of chiral molecules and thus facilitating chromatographic enantioseparation.

To achieve this goal, the CMRT dataset is established by automatically extracting experimental results from the relevant literature, the extraction procedure of which is provided in Supplementary Information S2.1. Several interesting aspects of the field of asymmetric catalysis can be reflected in the statistical analysis of the dataset including the contribution of authors, the average new enantiomers reported in the literature, and the usage frequency of HPLC columns, which are detailed in Supplementary Information S2.2. Meanwhile, visualization of molecules in the CMRT dataset is provided in Supplementary Information S2.3.

**Construction of QGeoGNN**

Benefiting from the natural graphic attribute of molecular structure, graph representation has attracted increasing attention in recent years[28]. The atoms and chemical bonds in the molecule are easy to be interpreted as a graph, which is referred to as Graph G (Fig. 2a). The node and edge features in Graph G are related to the characteristics of the molecular atoms and bonds, respectively. Meanwhile, considering that the bond length and angle can reflect the information of 3D conformation, another bond-angle graph, Graph H, is constructed as a complement for Graph G to incorporate geometry characteristics. In Graph H, the node feature is the bond length and the edge feature is the bond angle (Fig. 2a). Compared with traditional molecular representations, the graph representation can reflect charity by the chiral tags for labeling the handedness of chiral centers. Based on Graph G and H, the quantile geometry-enhanced graph neural network (QGeoGNN) is constructed. As illustrated in Fig. 2b, the proposed QGeoGNN takes experimental settings (i.e., elution proportion) into consideration, which makes the framework more appropriate to address practical experimental scenes. Meanwhile, the incorporation of relevant molecular descriptors further assists to distinguish enantiomers from macroscopic molecular properties. Through the graph convolution, the graph representations are obtained and then transformed into the prediction through a fully-connected layer. Details for the construction of QGeoGNN are provided in the Method section and Supplementary Information S1.

According to the chromatographic process equation[29], there exists an inverse proportional relationship between the retention time and flow rate, which is written as[29]:

$$RT = t_0(1 + K\frac{V_s}{V_m}) \approx \frac{1}{v}(V_m + KV_s), \qquad (1)$$

where $RT$ is the retention time, $K$ is the partition coefficient, $v$ is the flow rate, $V_m$ and $V_s$ are the volume of mobile and stationary phase, $t_0$ is the dead time respectively. This equation is verified by experiments in Supplementary Information S3.1 and the results confirm that it satisfies the requirements of this work. Therefore, the prediction target is set to be $RT \times v$ (abbreviated as $RT_v$) in this work to incorporate the chromatographic process equation. Considering that a) the dataset is collected from diversified articles which may be affected by objective factors, and b) the intrinsic floatability of RT, the quantile learning technique is innovatively incorporated into the deep learning model to take the data uncertainty into account. Quantile learning is originated from the quantile regression[30] which essentially regards the prediction target as a variable to learn its quantiles. Compared with the traditional least squares regression, quantile regression can describe the statistical distribution of variables in more detail. Inspired by the quantile regression, the quantile loss $L_\alpha$ is defined in deep learning so that the model can output quantiles of the predicted value, which is written as:

$$L_\alpha(y^{true}, y^\alpha) = \sum_{i=y_i^{true}<y_i^\alpha}(\alpha-1)|y_i^{true} - y_i^\alpha| + \sum_{i=y_i^{true}\geq y_i^\alpha}\alpha|y_i^{true} - y_i^\alpha|, \qquad (2)$$

where $\alpha$ is the quantile, $y^{true}$ and $y^\alpha$ are the observed data and the quantile prediction. In this work, the loss function of QGeoGNN consists of three parts, namely quantile loss, quantile limit, and deadtime limit (Fig. 2c). The quantile loss enables the QGeoGNN to learn the predicted value, 90th quantile, and 10th quantile simultaneously, while the quantile limit and deadtime limit function as the constraints to make outputs conform to the mathematical and physical restriction. Deep quantile learning especially fits the requirements of chromatographic enantioseparation in this work, since the experimental results usually vary within a certain range due to the influence of objective factors such as sample concentration, humidity, equipment conditions, and ambient temperature. It means that quantile learning is more conducive to eliminating the impact of objective errors, and thus facilitating the prediction of chromatographic enantioseparation.

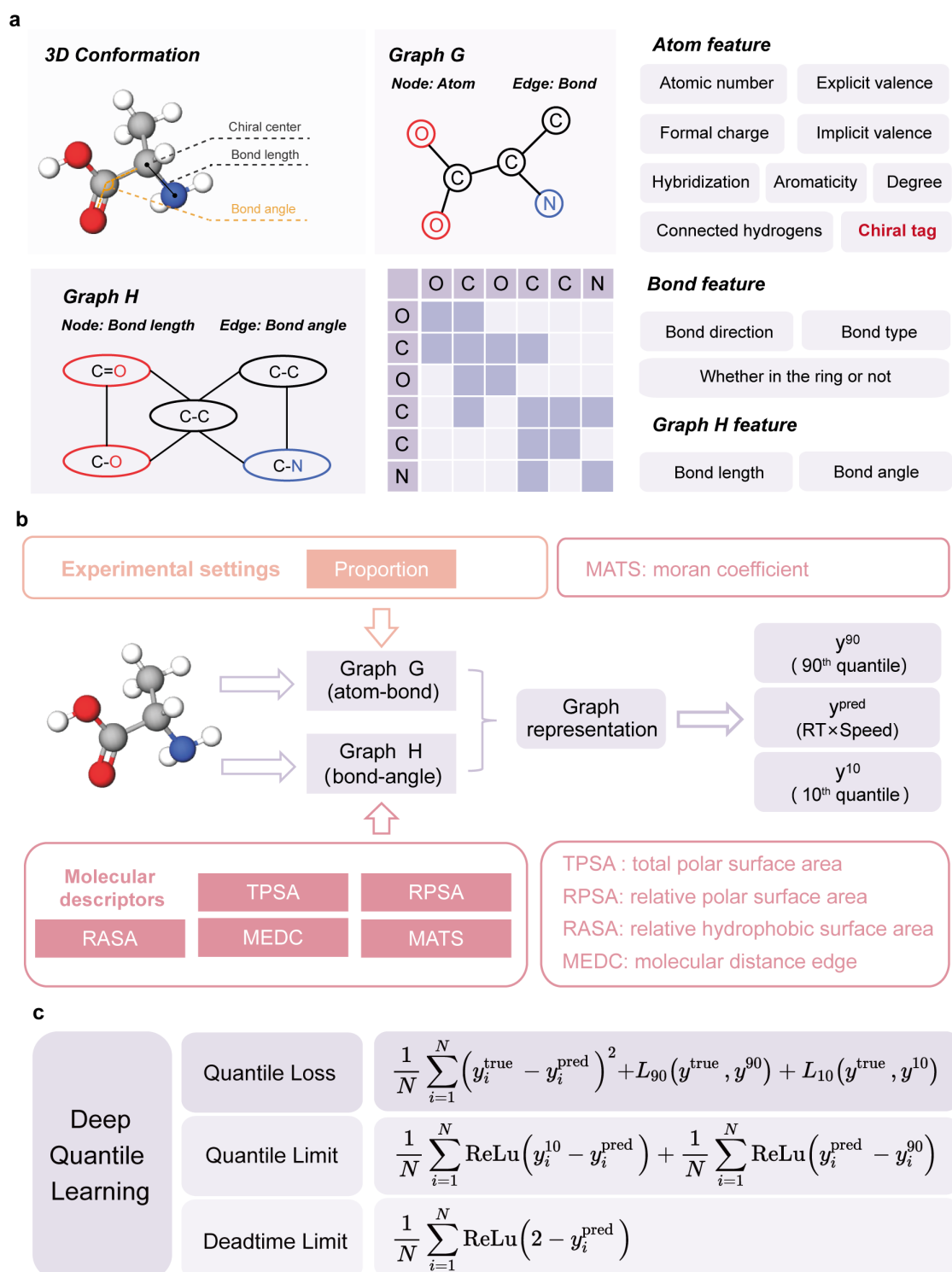

**Fig 2. The construction of QGeoGNN. a,** An example of the translation from the 3D conformation of a molecule to two graphs including the atom-bond graph G and the bond-angle graph H. Each graph is consisted of node, edge, and corresponding features. The edges are represented in the form of adjacent matrix. **b,** The scheme of the QGeoGNN, the experimental settings, and molecular descriptors are incorporated in Graph G and H, respectively. The output neuron of QGeoGNN is 3, namely, the 90$^{th}$ quantile, the prediction, and the 10$^{th}$ quantile. c, The constituent of the loss function in QGeoGNN, including the quantile loss, the quantile limit, and the deadtime limit.

**Single-column prediction**

As illustrated in Fig. 3a, in the area of asymmetric catalysis, various kinds of columns are adopted to handle diversified molecules due to the difficulty of chromatographic enantioseparation. The differences between HPLC column types come from many aspects like CSPs and column model, which affects the chiral recognition ability of HPLC column towards a wide variety of compounds. Among them, ADH, ODH, IC, IA, and OJH are the most frequently utilized column types in the dataset that we collected. From Fig. 3b, it can be seen that the probability density distribution of the retention times in these columns is similar, where the RTs of most molecules are in the range of 5 min to 30 min. To better demonstrate the prediction ability of the proposed QGeoGNN, in this section, single-column prediction is conducted where a prediction model is created in each column type. The benefits of the single-column prediction lie in that the conditions of the column are fixed in each predictive model, which means that the data has a good consistency and is conducive to learning the underlying molecule structure-retention time relationship. For each model, the dataset is split into the training dataset, validating dataset, and testing dataset by 90/5/5. The training dataset is utilized to train the model and the validating dataset is employed for early stopping. The testing data is utilized to examine the model's performance of the out-of-sample prediction. Considering the distribution of RT values, data points with $RT_v$ greater than 60 are dropped. The predicted results and corresponding root mean squared error (RMSE) and $R^2$ are shown in Fig. 3c. It is observed that the QGeoGNN has a good predictive ability for each column with $R^2$ all larger than 0.7, which indicates that the molecular structure-retention time relationship has been learned well. To eliminate the influence of randomness in splitting the dataset, 10 independent trials are conducted with different random seeds for each column type, and the results show that although the model is affected by randomness to a certain extent, it remains satisfactory prediction ability statistically. More details are provided in Supplementary Information S3.2.

To better reveal the predictive ability of the proposed QGeoGNN, the influence of data volume and noise is investigated where the prediction on the ODH column is taken as an example. The data noise is added as:

$$\hat{y} = y + \varepsilon \cdot std(y) \cdot N(0,1) \tag{3}$$

where $\hat{y}$ is the noisy data, $y$ is the observation data, $std(y)$ refers to the standard deviation of the observation data, and $N(0,1)$ is the normal distribution. The results are illustrated in Fig. 3d. It is discovered that the QGeoGNN is robust to data noise since the performance maintains stability faced with 10% data noise. At the same time, the observation data itself has inevitable experimental errors, which further verifies the superiority of the proposed method in dealing with noise. In terms of data volume, it is found that the prediction accuracy increases with the increase of training data ratio, and the trend of increase keeps apparent with 90% of the data, which means that if more sufficient data is provided, the prediction accuracy of QGeoGNN still has ample room for improvement.

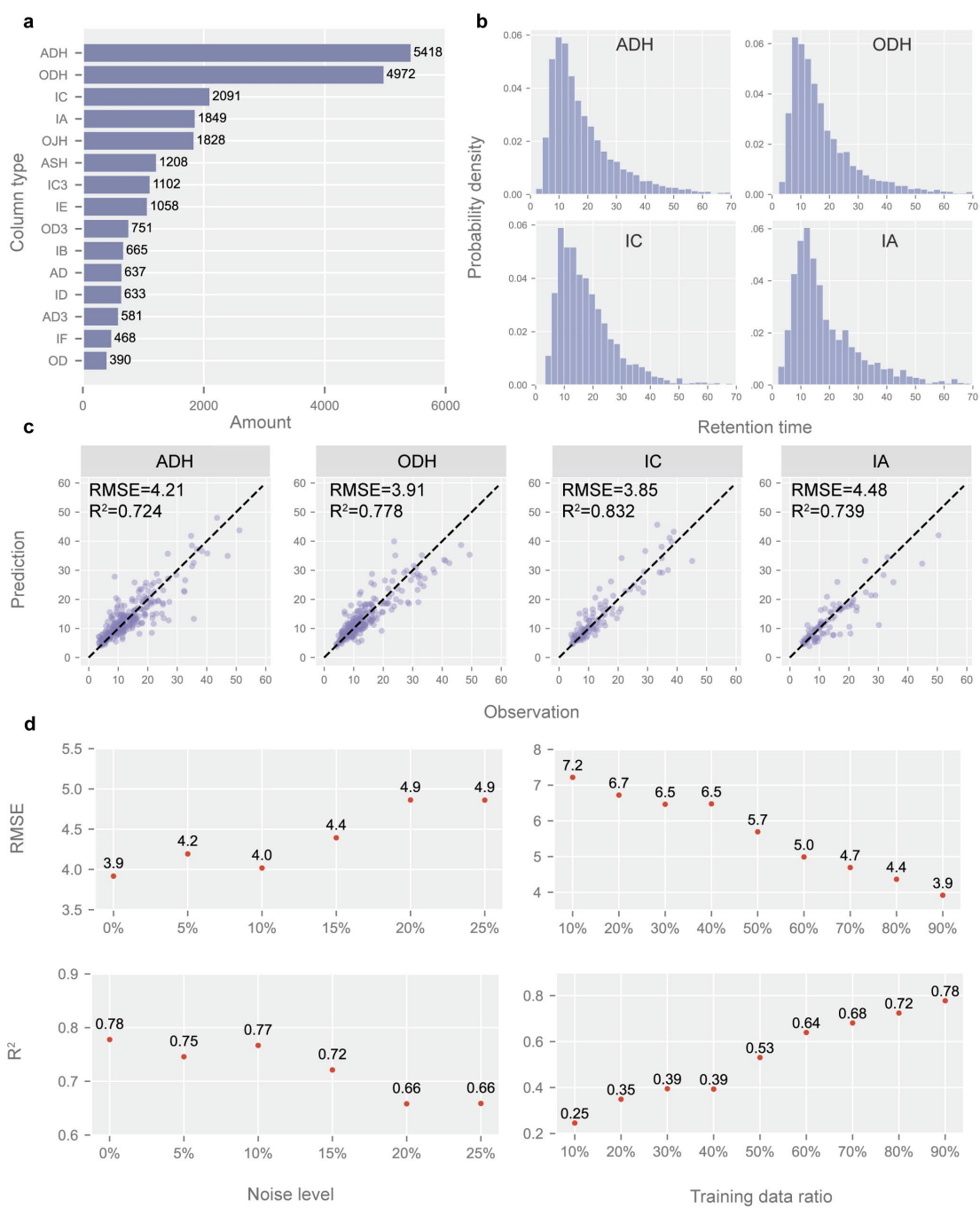

**Fig. 3. The performance of QGeoGNN for single-column prediction. a,** The data amount of each column type in the CMRT dataset established in this work. Only the first 15 column types with the largest volume are displayed here. **b,** The probability density distribution of the retention time in the subset of ADH, ODH, IC, and IA columns. **c,** Observation versus prediction for the proposed QGeoGNN to predict out-of-sample molecules. The dataset is randomly split into 90/5/5 by compounds, and only testing data are shown in plots. The dashed line is the $y = x$ line. **d,** the influence of data noise (left) and data volume (right) in ODH column data. The measurements are the root mean squared error (RMSE) and $R^2$.

**Multi-column prediction**

On the basis of the satisfactory performance of QGeoGNN in single-column prediction which confirms that the proposed framework is able to learn the molecular structure–retention relationship well, multi-column prediction is conducted in this section that acquires to integrate the prediction of diversified types of columns into a synthetic model. Here, the domain knowledge of chromatography is combined with machine learning techniques to facilitate model construction. In the HPLC column that is depicted in Fig. 4a, CSPs are derived from polysaccharides, including cellulose and amylose which are some of the most common chiral bio-based polymers in nature. As the insufficient chiral recognition ability of cellulose and amylose, their derivatives such as esters and carbamates modified with corresponding substituents are more frequently applied for both analytical and preparative enantioseparations[31]. The CSPs are usually immobilized or coated to silica gel. Therefore, in this work, three main factors are taken into consideration that affect the chiral recognition performance of HPLC columns, including the CSPs, the connection type (immobilized or coated), and the packing material (silica) size.

In this work, all types of HPLC columns in the dataset are composed of different collocations of two substrates and seven substituents (Fig. 4a). The connection type is encoded into one-hot code, which is incorporated into the edge features in Graph G of QGeoGNN along with the packing material size. The properties of CSPs are described by relevant descriptors and are added to the edge features in Graph H. In this way, all data in the CMRT dataset can be adapted to train a synthetic model for multi-column prediction, which enhances the availability of data. Considering that it is unrealistic to establish single-column prediction models for some less frequently used columns, where the data volume is small and insufficient for model construction, the multi-column prediction models combine the domain knowledge of chromatography with the machine learning model so that it can handle a variety of columns, which further improves the flexibility and scalability of the framework. The predictive performance of the multi-column prediction models is illustrated in Fig. 4b, where the entire data is split into 90/5/5 and the prediction on testing data is depicted in the figure. Faced with data from diversified columns and experimental conditions, the $R^2$ of the predictive model still achieves 0.702, which confirms the predictive ability of the synthetic model. An additional experiment is conducted in Supplementary Information S3.4 to observe the influence of column features in the multi-column prediction, and the results show that the incorporation of column information is of great importance for the accuracy of QGeoGNN.

In order to better demonstrate the superiority of the proposed QGeoGNN, conventional machine learning techniques, including LGB, XGB, artificial neural network (ANN), and GNN, are adopted to train prediction models for comparison. In LGB, XGB, and ANN, the molecular fingerprints and descriptors are employed for representation, while GNN only utilizes Graph G for molecular representation. The column information is incorporated into these models as well and other conditions are kept the same. More details of implements of these conventional models are provided in the Method section. The results are provided in Fig. 4b. It is obvious that classic tree-based models like LGB and XGB have poor predictive performance. On the other hand, the ANN can grasp the general relationship, but the accuracy needs to be improved. In comparison, the GNN-based models show superior ability where $R^2$ of GNN and QGeoGNN achieve 0.620 and 0.702, respectively. This is an interesting phenomenon since the conventional tree-based and ANN-based models often perform well in most chemical molecular prediction tasks, including previous literature on RT prediction[23–25]. In order to reveal the reasons behind this phenomenon explicitly, we provide the prediction of two pairs of example enantiomers given by different machine learning models in Fig. 4c. It is evident that the conventional models have difficulty in differentiating enantiomers since the predicted RTs are very close, even the same, which accounts for the poor performance on the CMRT dataset with enantiomers. In contrast, the QGeoGNN can not only distinguish enantiomers well but also provide an accurate predicted RT and its value range. The results demonstrate that the representation of chiral information is of great significance in the chromatographic enantioseparation task considered in this work, and the graph is proven to be a superior representation method than the conventional molecular fingerprints and descriptors when dealing with enantiomers. In addition, compared with GNN, the additional Graph H that incorporates the information of 3D conformation assists to learn the inherent molecular structure–retention relationship, which further improves the predictive ability of the model.

**Chromatographic enantioseparation probability assessment**

The ultimate goal of the retention time prediction model is facilitating chromatographic enantioseparation, which has been an outstanding issue all along. The difficulties for machine learning mainly concentrate on two aspects including chiral representation and error sensitivity. Specifically, chiral representation decides the ability to distinguish enantiomers while the error sensitivity determines the accuracy. Benefiting from the geometry-enhanced graph representation and the quantile learning, the proposed QGeoGNN provides a promising way to handle the above-mentioned challenges and thus facilitate chromatographic enantioseparation. In order to quantitatively evaluate the probability of enantioseparation under certain experimental conditions like column types, flow rate, and elution proportion, a chromatographic separation probability $S_p$ is defined in this work based on the predicted value ranges of enantiomers. The formula is written as:

$$S_p = 1 - \frac{L_{overlap}}{RT_{90}^{max} - RT_{10}^{min}}, \qquad (4)$$

where $L_{overlap}$ refers to the overlapping length of the predicted value ranges of enantiomers, $RT_{90}^{max}$ and $RT_{10}^{min}$ are the maximum of 90th percentile and the minimum of 10th percentiles of enantiomers, respectively (Fig. 5a). Some examples are given in Fig. 5a to better illustrate the separation probability. It is found that the proposition of $S_p$ assists to eliminate the impact of prediction errors to some extent. Specifically, single predicted values have low fault tolerance since the separation threshold is rigorous (usually tens of seconds). In comparison, quantile learning provides potential value ranges in consideration of data uncertainty, which can provide a separation probability instead of simple yes or no, which improves fault tolerance rate, and is more meaningful for chromatographic separation. Meanwhile, chromatographic enantioseparation prediction requires the model to learn the difference between enantiomers well. It means that although the error of the predicted $RT_v$ in multi-column prediction ($RMSE$=5.27) seems to be much greater than the separation threshold, chromatographic enantioseparation prediction is still possible.

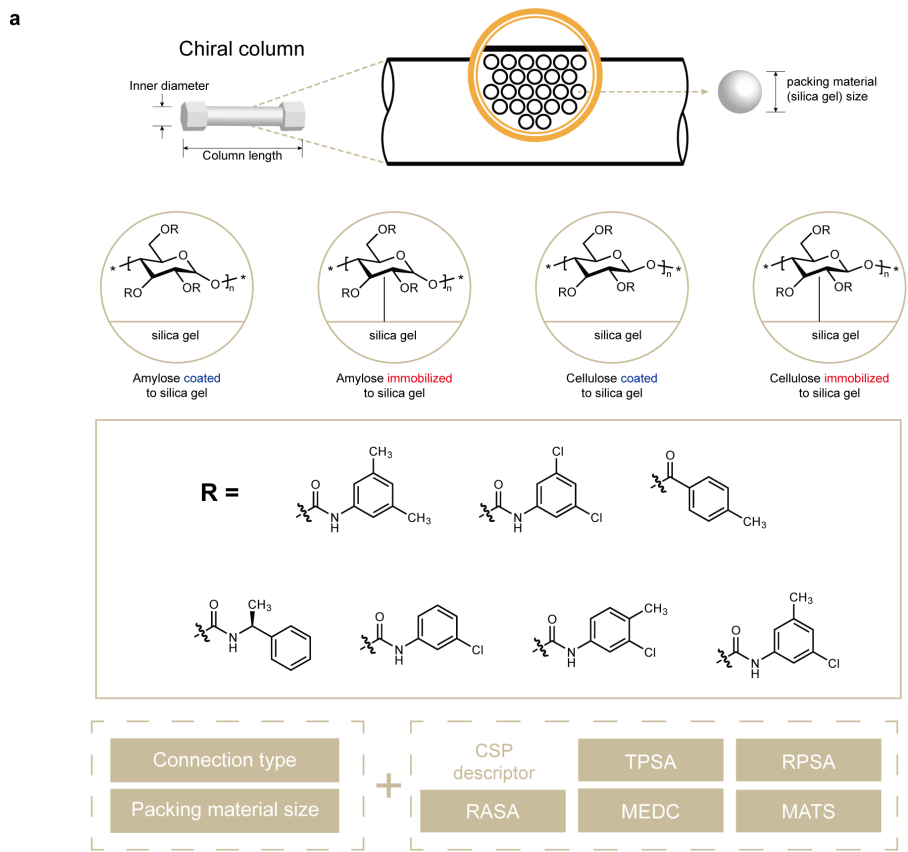

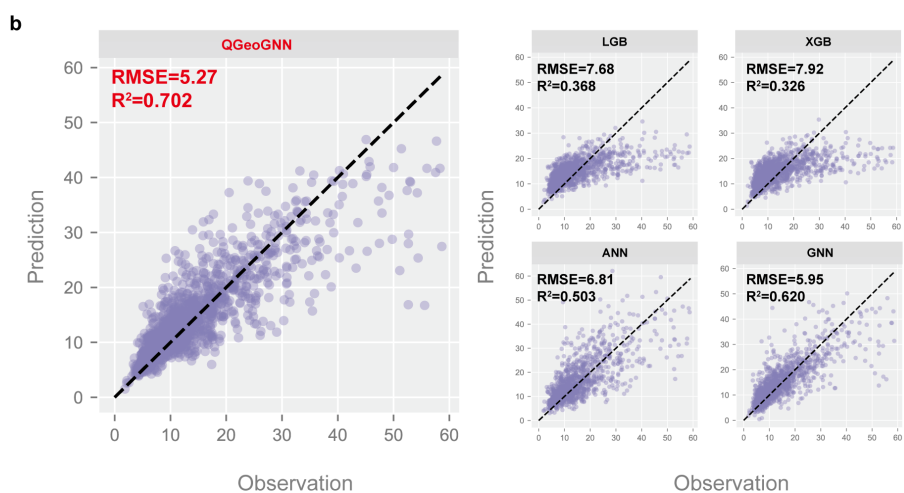

**Fig. 4. The comparison between QGeoGNN and conventional machine learning techniques for multi-column prediction. a,** The domain knowledge of chromatography by HPLC, which is consisted of the chiral stationary phase (CSP) and mobile phase. For CSPs, the packing material size, substrates, substituents, and connection type (immobilized or coated) will affect the chiral recognition ability of HPLC columns. The descriptors of CSPs, connection type, and packing material size are incorporated into the QGeoGNN for multi-column prediction. **b,** Observation versus prediction for the proposed QGeoGNN, LGB, XGB, artificial neural network (ANN) and graph neural network (GNN) to predict out-of-sample molecules in diversified HPLC columns. The entire dataset is randomly split into 90/5/5, and only testing data are shown in plots. The dashed line is the $y = x$ line. **c,** Predicted retention time of two pairs of example enantiomers given by different machine learning models. The result of QGeoGNN is presented in the form of the 10[th] quantile~90[th] quantile (predicted RT).

In order to demonstrate the ability of the proposed model to facilitate chromatographic enantioseparation, several experiments are conducted in this section. First, 412 pairs of enantiomers (i.e., 824 data) are randomly selected from the CMRT dataset to form the testing dataset while the training dataset (23,020 data) and validating dataset (904 data) are randomly chosen from the remaining data to train the prediction model. The separation probability $S_p$ of the enantiomers in the testing dataset is calculated and illustrated in Fig. 5b. In this work, the separation probability is divided into five gears, including extremely low (<0.2), low (0.2~0.4), medium (0.4~0.6), high (0.6~0.8), extremely high (0.8~1.0). In order to calculate the accuracy of separation prediction, we regard those enantiomers with $S_p$>0.4 as separable, and the accuracy reaches 84.7%. Meanwhile, the accuracy with different thresholds of $S_p$ is provided in Fig. 5c. The high accuracy confirms the ability of the proposed QGeoGNN to predict chromatographic enantioseparation. It is worth noting that chromatographic enantioseparation prediction differs from typical classification tasks owing to its distinctive requirements. Considering the difficulty of chromatographic enantioseparation, a suitable separation condition is very important but scarce, which means that it is unaffordable to predict an actually separable situation as inseparable (Type I error) since it may miss the precious suitable separative conditions. In contrast, it is relatively acceptable to predict the inseparable condition as separable (Type II error), because this will only induce additional experiments. Benefiting from the multi-column prediction accomplished by QGeoGNN, the separation probabilities of the same enantiomers in different types of HPLC columns can be obtained and compared, which can directly reflect the predicted chiral recognition ability of each column to the given enantiomers, thus providing suggestions on selecting proper experimental conditions without tedious trails and errors. In Fig. 5d, we provide an example of the utilization in practical application. To separate enantiomers, multiple candidate conditions composed of six column types and corresponding proportions and flow rates are considered to select the most proper separative condition. If experiments are conducted to try with all these conditions, it will take several hours. In contrast, the QGeoGNN only needs seconds to predict the separation probability under each condition, which can be visually depicted in the figure, and easy to find the most proper situations with the largest $S_p$ and moderate predicted retention time, thus saving appreciable time for experimenters. The experimental experiments confirm that the enantiomers can only be separated in the IG column, which is consistent with the prediction.

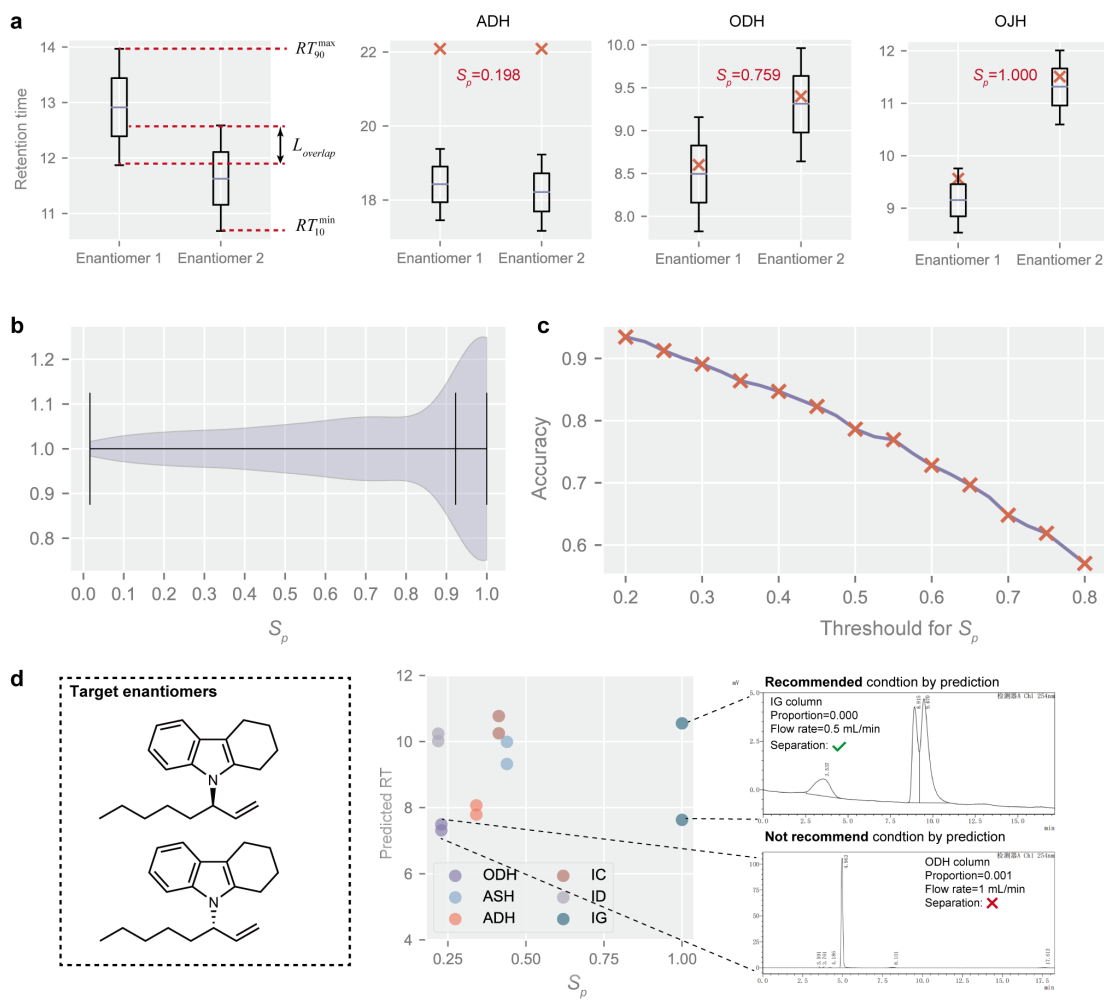

**Fig. 5. Definition and application of Chromatographic enantioseparation probability assessment. a,** The definition of separation probability and some examples under different conditions, including an inseparable situation under ADH column and two separate situations under ODH and OJH columns. The red cross refers to the observed data while the error bar refers to the predicted value range (10$^{th}$ quantile~90$^{th}$ quantile). The purple line in the error bar is the predicted RT. **b,** The violin plot of the distribution of calculated $S_p$ for 412 pairs of testing enantiomers. The black vertical lines mean the minimum, median, and maximum values, respectively. **c,** The accuracy with different thresholds of $S_p$ (i.e., regard $S_p$>threshold to be separable). **d,** An example of the utilization in practical application, including the enantiomers (left), prediction of different columns made by multi-column prediction model (middle), and verification experiments (right).

## Discussion

In this work, a new research framework is proposed to incorporate machine learning techniques into the field of experimental chemistry to promote the efficiency of the researchers practically when faced with chromatographic enantioseparation. The proposed framework of quantile geometry-enhanced graph neural network (QGeoGNN) focuses on addressing several core issues including data acquisition, chiral molecular 3D representations, and data uncertainty. In most conventional machine learning research, the data can be obtained easily by numerical simulation or scientific computing. However, data acquisition in experimental science is always scarce due to expensive and time-consuming experiments. For example, each HPLC experiment considered in this work may take tens of minutes. It means that merely relying on conducting experiments manually to collect data is unaffordable. Benefiting from the consistency of standardized commercial HPLC columns, this work manages to construct a precious chiral molecular retention time dataset from numerous articles in the area of asymmetric catalysis and thus solve the problem of data acquisition.

On the other hand, a specialized graph neural network called QGeoGNN is established by incorporating molecular 3D conformation, experimental conditions, relevant descriptors, and quantile learning to be more suitable for experimental practice. Experiments have confirmed that the QGeoGNN has a satisfactory ability to predict the retention time of chiral molecules in single-column prediction. Furthermore, we combine the domain knowledge of chromatography with machine learning techniques to be closer to the practical demand. The factors that affect the chiral recognition ability of the HPLC column are summarized and characterized, which are incorporated into the machine learning model to accomplish multi-column prediction. Experiments show that the QGeoGNN is able to predict the *RT* values under diversified conditions, which paves the way for practical application.

Finally, the ability of the proposed framework to facilitate chromatographic enantioseparation is examined. The experimental results are usually affected by objective errors, and thus inherently have uncertainty. The quantile learning technique utilized in this work attempts to capture the uncertainty during the training process and can provide a value range. On this basis, a novel concept called separation probability is defined in this work to measure the predictive probability of enantiomers being separated under a given condition. As a consequence, the QeoGNN can predict the separation probability in diversified conditions quickly and flexibly, and recommend suitable conditions by comparison, which will promote the efficiency of chromatographic enantioseparation.

At present, this research remains some shortcomings that can be improved in the future. First, the representativeness and quality of the data are uncontrolled and sometimes biased since they are extracted from existing literature, which will affect the predictive performance of the machine learning model. Second, prediction accuracy still needs to be improved when faced with completely unfamiliar molecules, which is detailed in Supplementary Information S3.5. Third, the feature extraction is worth further investigating to better represent chiral-related information, since the representations of molecules and HPLC columns are closely related to the prediction accuracy of QGeoGNN. More comprehensive discussions about the advantages and inadequacies of this research are provided in Supplementary Information S3.6. Despite these limitations, we believe this framework can be directly put into practical use to help experimenters save time and improve efficiency by avoiding tedious attempts to determine proper experimental conditions.

## Methods

**The graph representation for molecules.** In this work, the molecular representation is accomplished by constructing two graphs, Graph G and H (Fig. 2a). In data science, graphs are often used to describe unstructured data like social networks, chemical molecules, and traffic networks. A typical graph is composed of several nodes and edges that indicate the connection relationship. As illustrated in Fig. 1a, Graph G expresses the plane structure of the chemical molecule where the nodes refer to the atoms and the edges refer to the chemical bonds. The feature of each node in Graph G contains 9 properties of the corresponding atom, including the atomic number, chiral tag, degree, explicit valence, formal charge, hybridization, implicit valence, aromaticity, and the number of connected hydrogen atoms. The feature of each edge in Graph G contains 3 properties of the corresponding bond including bond direction, bond type, and whether in the ring or not. In this work, the experimental condition, elution proportion, is also added to the feature of each bond. On the other hand, Graph H describes the 3D conformation of the molecule where nodes refer to the bond and edges refer to the bond angle (Fig. 2a). The feature of each node in Graph H contains only one property, the length of the corresponding bond, while the feature of each edge in Graph H contains six properties including the bond angle and five relevant descriptors, namely total polar surface area (TPSA), relative polar surface area (RPSA), relative hydrophobic surface area (RASA), molecular distance edge (MEDC), and moran coefficient (MATS). The molecular descriptors are calculated by the python package *Mordred* and chosen according to the spearman coefficient that identifies the correlation with retention time. In multi-column prediction, the features of HPLC columns are incorporated in the edge feature of Graph G. More details about the graph representation are provided in Supplementary Information S1.

**Quantile geometry-enhanced graph neural network (QGeoGNN).** The QeoGNN is constructed based on the graph representation and the graph isomorphism network (GIN). The fundament of GIN is the graph isomorphic convolution layer (GINConv)[32], which is defined as:

$$x'_i = h_\theta((1+\varepsilon) \cdot x_i + \sum_{j \in N(i)} x_j),$$

where $x'_i$ and $x_i$ are the node representations in the next layer and current layer, respectively. $x_j$ is the representation in the adjacent nodes. $h_\theta$ is a multilayer perceptron (MLP) and $\varepsilon$ is a constant that equals 0 in this work. In the QGeoGNN, the node embedding is performed for each node in Graph H to obtain the corresponding node representation based on GINConv. Then, the node representation of Graph H is added to the edge representation of Graph G to build a bridge for information transmission between Graph G and Graph H. Afterwards, the node representation of each node in Graph G obtained by node embedding is pooled to get the graph representation. Finally, a fully-connected layer is used to transform the graph representation into the prediction. Deep quantile learning is incorporated into the QGeoGNN by modifying the loss function. As illustrated in Fig. 2c, the quantile loss, quantile limit, and deadtime limit work together to enable the QGeoGNN to learn the value range. Therefore, the input of QGeoGNN is two molecular graphs of each molecule that are converted from the 3D conformation, and the output neuron is 3, including the 90[th] quantile, the predicted value, and the 10[th] quantile. More details about the construction of QGeoGNN are provided in Supplementary Information S1.

**Experimental settings and parameters.** In the QGeoGNN utilized in this work, the number of GINConv is 5, the graph pooling strategy is the summation, the embedding dimension of the node and edge representation is 128, and the batch size is 2,048. The training epoch is 1,500, and the validate loss is adapted for early stopping. The optimizer is Adam and the learning rate is 0.001. For single-column prediction, the prediction models are established for ADH, ODH, IC, and IA columns, respectively. For each column, the sub-dataset is randomly divided into 90/5/5 to obtain the training, validating, and testing dataset. For multi-column prediction, the entire dataset is split into 90/5/5 to train a synthetic model. For comparison, the XGB, LGB, ANN, and GNN are also employed to train a predictive model. The input of XGB, LGB, and ANN is composed of the 167-dimensional

MACCS keys that are utilized to represent the molecular structure, the 5-dimensional molecular descriptors that are the same as those utilized in QGeoGNN, and 3-dimensional column information. For XGB, the number of estimators is 200, the maximum depth is 3, and the learning rate is chosen to be 0.1. For LGB, the maximum depth is 5, the learning rate is 0.007, the number of leaves is 25, and the number of estimators is 1000. For ANN, there are 3 hidden layers with 50 hidden neurons in each hidden layer. The activation function is leaky ReLu and the optimizer is Adam with a learning rate of 0.001. The training epoch is 10,000 and early stopping is adopted. The construction of GNN is similar to QGeoGNN while it only has Graph G, and the loss function is of the mean squared error between the predicted and observed value. The column information is incorporated into the edge features for GNN.

**Data availability**
The CMRT dataset will be publicly available as of the date of publication.
All original code will be publicly available as of the date of publication.

**Acknowledgements**
We thank Prof. Changkun Li at Shanghai Jiaotong univeristy for invaluable discussions and helpful suggestions. This work is supported by the Natural Science Foundation of China (Grant Nos. 22071004, 21933001, 22150013).

**Author contributions**
F.M. constructed the chiral molecular retention time (CMRT) dataset. J.L. and H.X. analyzed the data. H.X. performed chemoinformatic and machine learning studies. H.X. and F.M. wrote the manuscript. F.M. and D.Z. supervised the whole project.

**Competing interests**
The authors declare no competing interests.

**References**
1. Artrith, N. *et al.* Best practices in machine learning for chemistry. *Nat. Chem.* **13**, 505–508 (2021).
2. Muratov, E. N. *et al.* QSAR without borders. *Chemical Society Reviews* vol. 49 3525–3564 (2020).
3. Gupta, R. *et al.* Artificial intelligence to deep learning: machine intelligence approach for drug discovery. *Mol. Divers.* **25**, (2021).
4. Ishida, S., Terayama, K., Kojima, R., Takasu, K. & Okuno, Y. Prediction and Interpretable Visualization of Retrosynthetic Reactions Using Graph Convolutional Networks. *J. Chem. Inf. Model.* **59**, (2019).
5. Coley, C. W., Green, W. H. & Jensen, K. F. Machine Learning in Computer-Aided Synthesis Planning. *Acc. Chem. Res.* **51**, (2018).
6. Shen, Y. *et al.* Automation and computer-assisted planning for chemical synthesis. *Nat. Rev. Methods Prim.* **1**, (2021).
7. Tkatchenko, A. Machine learning for chemical discovery. *Nature Communications.* **11** (2020).

8. Dobbelaere, M. R., Plehiers, P. P., Van de Vijver, R., Stevens, C. V. & Van Geem, K. M. Machine learning in chemical engineering: strengths, weaknesses, opportunities, and threats. *Engineering* **7**, (2021).
9. Haghighatlari, M. *et al.* Learning to Make Chemical Predictions: The interplay of feature representation, data, and Machine Learning Methods. *Chem* **6** (2020).
10. Burger, B. *et al.* A mobile robotic chemist. *Nature* **583**, 237–241 (2020).
11. Xu, H. *et al.* High-throughput discovery of chemical structure-polarity relationships combining automation and machine-learning techniques. *Chem* 1–13 (2022) doi:10.1016/j.chempr.2022.08.008.
12. Weininger, D. SMILES, a Chemical language and information system: 1: introduction to methodology and encoding rules. *J. Chem. Inf. Comput. Sci.* **28**, 31–36 (1988).
13. Rogers, D. & Hahn, M. Extended-connectivity fingerprints. *J. Chem. Inf. Model.* **50**, (2010).
14. Moriwaki, H., Tian, Y. S., Kawashita, N. & Takagi, T. Mordred: A molecular descriptor calculator. *J. Cheminform.* **10**, (2018).
15. Fang, X. *et al.* Geometry-enhanced molecular representation learning for property prediction. *Nat. Mach. Intell.* **4**, 127–134 (2022).
16. Zhou, G. *et al.* Uni-Mol: A universal 3D molecular representation learning framework. Preprint at https://chemrxiv.org/engage/chemrxiv/article-details/6318b529bada388485bc8361 (2022).
17. Blaser, H. U. Chirality and its implications for the pharmaceutical industry. *Rend. Lincei* **24**, (2013).
18. Brandt, J. R., Salerno, F. & Fuchter, M. J. The added value of small-molecule chirality in technological applications. *Nature Reviews Chemistry* **1** (2017).
19. Peluso, P. & Chankvetadze, B. Recognition in the domain of molecular chirality: from noncovalent interactions to separation of enantiomers. *Chem. Rev.* (2022) doi:10.1021/acs.chemrev.1c00846.
20. Okamoto, Y. & Ikai, T. Chiral HPLC for efficient resolution of enantiomers. *Chem. Soc. Rev.* **37**, 2593–2608 (2008).
21. Sun, L. *et al.* A simple method for HPLC retention time prediction: Linear calibration using two reference substances. *Chinese Med. (United Kingdom)* **12**, (2017).
22. Usman, A. G., Işik, S. & Abba, S. I. A novel multi-model data-driven ensemble technique for the prediction of retention factor in HPLC Method Development. *Chromatographia* **83**, 933–945 (2020).
23. Osipenko, S. *et al.* Machine learning to predict retention time of small molecules in nano-HPLC. *Anal. Bioanal. Chem.* **412**, 7767–7776 (2020).
24. Domingo-Almenara, X. *et al.* The METLIN small molecule dataset for machine learning-based retention time prediction. *Nat. Commun.* **10**, 1–9 (2019).
25. Low, D. Y. *et al.* Data sharing in PredRet for accurate prediction of retention time: Application to plant food bioactive compounds. *Food Chem.* **357**, (2021).
26. Eriksson, T., Björkman, S. & Höglund, P. Clinical pharmacology of thalidomide. *Eur. J. Clin. Pharmacol.* **57**, 365–376 (2001).
27. Francotte, E. R. Enantioselective chromatography as a powerful alternative for the preparation of drug enantiomers. *Journal of Chromatography A* vol. 906 379–397 (2001).
28. Jiang, D. *et al.* Could graph neural networks learn better molecular representation for drug discovery? A comparison study of descriptor-based and graph-based models. *J. Cheminform.*

**13**, 1–23 (2021).

29. Nicoud, R. M. Chromatographic Processes. *Cambridge University Press* (2015).
30. Das, K., Krzywinski, M. & Altman, N. Quantile regression. *Nat. Methods* **16**, 451–452 (2019).
31. Shen, J. & Okamoto, Y. Efficient separation of enantiomers using stereoregular chiral polymers. *Chem. Rev.* **116**, 1094–1138 (2016).
32. Xu, K., Jegelka, S., Hu, W. & Leskovec, J. How powerful are graph neural networks? *7th Int. Conf. Learn. Represent. ICLR 2019* 1–17 (2019).